\documentclass[useAMS,a4paper]{mn2e}
\usepackage{epsfig}

\newcommand{\mnras}{MNRAS}

\newcommand{\apj}{ApJ}

\newcommand{\apjs}{ApS}
\newcommand{\aj}{AJ}

\def\msun{{\rm M_{\odot}}}

\title [Disc clearing in M stars]
{The clearing of discs around late type T Tauri stars: constraints from
the infrared two colour plane}
\author[]{Barbara Ercolano$^{1,2,3}$,  Cathie J. Clarke$^{2}$ and
Alex C. Hall$^{2}$ \\
$^1$School of Physics, University of Exeter, Stocker Road, Exeter EX4 4QL\\
$^2$Institute of Astronomy, Madingley Rd, Cambridge, CB3 0HA, UK \\
$^3$Department of Physics and Astronomy, University College London, Gower Street, London, WC1E 6BT}

\date{Submitted: }

\pagerange{\pageref{firstpage}--\pageref{lastpage}} \pubyear{2003}

\begin{document}
\def\lta{\mathrel{\spose{\lower 3pt\hbox{$\mathchar"218$}}
     \raise 2.0pt\hbox{$\mathchar"13C$}}}
\def\gta{\mathrel{\spose{\lower 3pt\hbox{$\mathchar"218$}}
     \raise 2.0pt\hbox{$\mathchar"13E$}}}
\def\Msun{{\rm M}_\odot}
\def\msun{{\rm M}_\odot}
\def\Rsun{{\rm R}_\odot}
\def\Lsun{{\rm L}_\odot}
\def\19{GRS~1915+105}
\label{firstpage}
\maketitle

\begin{abstract}

We have undertaken SED modeling of discs around low mass T Tauri stars
in order to delineate regions of the infrared two colour plane (K - [8] versus 
K - [24]) that correspond to discs in different evolutionary stages. This
provides a ready tool for  classifying the nature of star-disc
systems based on infrared photometry. In particular
we demonstrate  
the distinct loci followed by discs that undergo
`uniform draining' (reduction in surface density by a spatially
uniform factor) from those that clear from the inside out. We  draw attention 
to the absence of objects on this `draining locus' in those
star forming regions  where the $24 \mu$m sensitivity would permit their
detection, as compared with  
the  $\sim 20$ objects in these regions with colours
suggestive of inner holes.  We thus conclude that discs predominantly
clear from the inside out.  We also apply our classification of the infrared two
colour plane to stars of spectral type M3-M5 in the IC 348 cluster and show that  some
of its members (dubbed `weak excess' sources by Muzerolle et al 2010)
that had previously been assumed to be in a state of clearing are instead
likely to be optically thick discs in which the dust is well settled towards
the mid-plane. Nevertheless, there are many discs in a state of partial
clearing in IC 348, with their abundance (relative to the total
population of disc bearing stars) being around four times higher than
for comparable stars in Taurus. However, the  number of partially
cleared discs relative to the total number of late type stars is similarly
low in both regions ($10$ and $20 \%$ respectively). We therefore conclude
that IC 348 represents a more evolved version of the Taurus population
(with more of its discs being highly settled  or  partially cleared)
but that the timescale for clearing is similarly short (a few times
$10^5$ years) in both cases.
\end{abstract}


\section{Introduction}
  The infrared colours of discs around young stars provide a prime
diagnostic tool for analysing the evolution of circumstellar discs.
Just as the distribution of stars on a Hertzsprung Russell diagram
contains information about the relative duration of various stellar
evolutionary stages, so infrared two colour diagrams have played
a similar role in the case of disc studies. In particular, Kenyon
\& Hartmann (1995) drew attention, in the case of young stars in the
Taurus star forming region, to a pronounced gap in this diagram,
intermediate between the colours of optically thick discs and those
of stellar photospheres. This suggested that young stars undergo
a {\it rapid} transition between disc possessing and discless status,
which is a small fraction (around $10 \%$) of their previous lifetimes
as disc bearing sources. Such `two timescale' behaviour has provided
a strong constraint on viable models of disc clearing, requiring
scenarios, such as photoevaporation (Clarke et al 2001, Alexander et
al 2006, Ercolano et al 2008, Ercolano, Clarke \& Drake 2009, Gorti \&
Hollenbach 2009, Owen et al 2010) or possibly planet formation
(Armitage \& Hansen 1999) where there is a final {\it rapid} clearing
phase. Discriminating between these two scenarios may become possible
in the future when disc statistics in clusters of different
metallicities start becoming available (Ercolano \& Clarke 2010, Yasui
et al 2009).

  In recent years, data acquired with the Spitzer Space Telescope has
permitted collation of infrared colours for stars in a range of 
star forming regions (Allen et al 2007; Evans et al 2009; Gutermuth et al 2009; 
Koenig et al 2008; Muench et al 2007; Rebull et al 2010). 
These provide samples that are, in some cases,
somewhat older than Taurus and which generally contain a higher proportion
of stars of later spectral type (M stars). Both these factors have been
invoked when explaining cases where the distribution of different
categories of infrared SEDs is rather different from that in  Taurus.

  A case where differences from Taurus have been the subject of
much debate is the cluster IC 348, which was initially age-dated at
roughly 2.5~Myr (Haisch et al 2001), but has since been revised to
4-5~Myr by Mayne et al (2007) and Mayne \& Naylor (2008). Spitzer
photometry has been acquired for this cluster by Lada et al (2006) and
Muench et al (2007)  with revised photometry and extra $24 \mu$m data
being added by Currie \& Kenyon (2009). Here Lada et al (2006) drew
attention to a population of what they called `anaemic' discs, being
objects that exhibited 
discs that were weak compared with those typical of Taurus. This
category includes sources which have alternatively been dubbed as 
`homologously depleted'
discs by Currie \& Kenyon (2009) and as three separate categories
by Muzerolle et al (2010) (i.e. `weak excess', `warm excess' and `classical
transition' discs). 
Each of these designations is designed to suggest discs that are
in a state of partial clearing. The number of such sources is large
in IC 348 (e.g. including sources where only an upper limit is
available at [24]~$\mu$m Lada et al (2006) classified 44 M3 to M5 stars in
IC 348 as having anaemic discs, compared with 41 with optically thick
discs; if only sources with a detection at [24]~$\mu$m are included then
the number of anaemic and optically thick discs are 18 and 31,
respectively). 

  The large number of discs that are apparently in a state of partial clearing
has prompted us (Section 2)
to investigate the expected trajectories in the
infrared two colour plane for M star discs that clear according to a variety
of scenarios. Our aim is therefore to deduce the likely physical state
of discs from their infrared colours and to provide a framework that can
be applied to the M stars in a range of clusters. We will avoid empirical
classifications (e.g. those that relate infrared excesses to the distribution
found in Taurus) and instead focus on the likely physical properties of the
discs concerned. We will however discuss how our classification relates to
the various empirical categories
of cleared discs mentioned above. 
 
  In Section 3 we analyse the case of IC 348 by assessing how the distribution
of sources in the infrared two colour plane can be mapped onto a distribution of discs in different evolutionary states. 
We will show that - although some of the
anaemic discs in this cluster are merely flat, optically thick discs with
no evidence for clearing (Luhman et al 2010)  
- there are nevertheless more partially cleared
discs among M stars in the intermediate age IC 348 cluster than in the case of M stars in the younger Taurus. Section 4 summarises our conclusions.

\section{ The evolution of M star discs in the infrared two colour plane}

\subsection{Model evolutionary tracks}
 We consider a range of disc clearing scenarios and plot their
trajectories in the K - [8]  versus K - [24] plane. K band emission
is largely contributed by the star whereas that at $8 \mu$m 
and $24 \mu$m originates in a broad radial range in the disc (roughly 0.2-0.3 AU
and 1-2 AU respectively for M-stars). Our choice of these bands is motivated
by the availability of data acquired by IRAC and MIPS respectively and
also by the fact that they are well separated in wavelength. As
pointed out by Ercolano, Clarke \& Robitaille (2009), the shorter wavelength infrared
bands are less useful for the study of disc clearing in M stars since
they can be dominated by the star even in the case of an optically
thick (uncleared) disc. 

\subsubsection {The case of zero thickness  discs}

\begin{figure}
\begin{center}
\includegraphics[width=8cm]{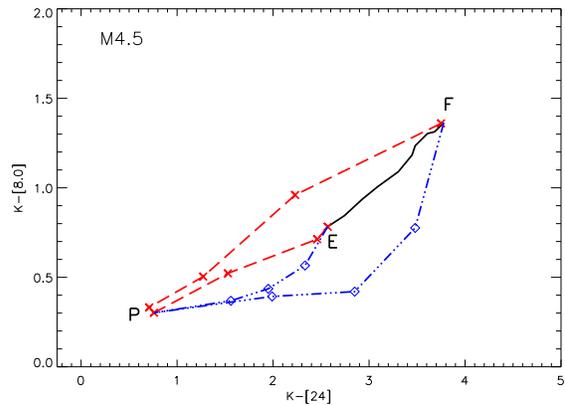}
\caption[]{Model evolutionary tracks for a zero thickness disc
  surrounding an M4.5 star. Point
  F is for an optically thick reprocessing disc seen almost face on
  (cos i = 0.95), point E is for the same disc seen almost edge on
  (cos i = 0.05) and  intermediate inclinations lie along the line
  FE. Point P represents the colours of a pure stellar photosphere.
 The red dashed lines show the
  evolution for uniform draining of the disc if seen close to face on (line
  FP) or close to edge on (line EP). The red 'X' mark nominal  ages 
  along the FP and
  EP  lines (1 Myr, 100 Myr, 469 Myr and 2.1 Gyr) based on an
 extrapolation of the disc's  viscous draining according to a 
 similarity solution (see text). The blue
  dash-dot-dot lines FP and EP show the evolution of a disc that is
  clearing via inside-out erosion. The three intermediate points
  marked by diamonds are for inner holes of [8], 20 and 40~AU.}
\label{f:f1}
\end{center}
\end{figure}

  We first consider the limiting case of razor thin, flat, optically
thick reprocessing discs, viewed at a range of inclination angles.
We then consider
two scenarios for disc clearing: a) progressive uniform reduction of the
disc surface density at all radii  and b) the carving out of the optically
thick disc by an inner cavity of progressively larger radius. {\footnote
{Note that
we use the term `uniform draining' for scenario a); in essence this is
the same situation that is termed `homologously depletion' by Currie et al.
We however use a separate term in order that we can use the label
`homologously depleted' for those sources {\it classified as such}
by Currie et al: these  do not necessarily lie on our model tracks for
uniform draining.}} In each case we compute the SED using the
radiative transfer code of Whitney et al (2003a,b).  

  Figure 1  shows the result of this exercise in the case of a star of
spectral type M4.5. The point F represents the colours
of a flat optically thick reprocessing disc that is almost
face on (cos i $=0.95$)  and the point E of
a similar disc viewed almost edge on (cos $i= 0.05$). The slope
of the line EF is consistent with the fact that tilting the
disc mainly results in a reduction of the $8$ and $24 \mu$m flux
rather than the (mainly stellar) emission at K. We would expect that
a disc with negligible accretion luminosity and where the dust was completely
settled to the mid-plane would, depending on its inclination, lie
on
the line FE. 

  The dashed line FP represents the trajectory of a face-on disc that
undergoes progressive uniform reduction in its surface density until
it arrives at P which has the colour of a pure stellar photosphere. 
Once the disc starts to become optically thin, its radial surface
density profile affects the SED: the model shown is for the case that
the surface density profile $\Sigma$ scales as $R^{-1}$.
The line EP is the corresponding locus for a nearly edge on disc. 
The slope of these lines reflects the fact that as the disc surface
density is reduced it is the longer wavelength emission (i.e. at
$24 \mu$m) that becomes optically thin first. We emphasise that very
small surface densities are required in order to achieve unit
optical depth in the mid infrared (i.e roughly 6$\times$10$^{22}
cm^{-2}$ at $8 \mu$m and 10$^{23} cm^{-2}$ at $24 \mu$m)   
and that the progression along FP or EP would be slow unless the
disc draining were accelerated. 

 In order to illustrate this last point
we place ticks on the line FP which correspond to nominal times in the
case of a viscously draining disc. This disc model has been normalised
so that we start from a disc  
 at an age
of 1Myr for which  the accretion rate on to the star has fallen to the point of marginal
detectability (i.e. $10 ^{-10} M_\odot$ yr $^{-1}$). The forward evolution
employs a viscous similarity
solution for a disc with steady surface density profile $\propto R^{-1}$
for which the disc surface density declines as $t^{-1.5}$ (Hartmann
et al 1998); the ticks correspond to ages $1 Myr$ (at F) and then
$100$ and $460$ Myr and finally $2.1$ Gyr by the time such a disc
would arrive at P. Given that many young stars already exhibit purely
photospheric
emission  at an age of a few Myr it is immediately obvious that
the later stages of disc clearing {\it cannot} be a simple extrapolation
of the viscous draining that dominates the earlier phases of disc evolution. 
If the lines FP and EP had any relevance to disc clearing they would
instead have to correspond to a faster depletion mechanism,
possibly related to the reduction in opacity due to grain growth or
a lowering of the  dust to gas ratio in the disc.

 We now instead consider the alternative scenario in which a disc that
has settled to a flat optically thick configuration (and thus, according
to its inclination, resides along the line FE) is now carved out from
the inside by the creation of successively larger inner holes. We draw
a couple of loci (dash-triple-dot lines in Figure~1) 
which originate at F and E and where the diamonds
mark hole  sizes of $8,20$ and $40$ AU.
These loci initially descend nearly vertically
as the creation of a small hole reduces the $8 \mu$m flux without
appreciably changing the $24 \mu$m flux; for hole sizes of around
$10$ AU  and larger,
the $24 \mu$m flux starts also to decline and the locus flattens out so
as to approach P nearly horizontally. 

\subsubsection{Finite thickness discs}

\begin{figure}
\begin{center}
\includegraphics[width=8cm]{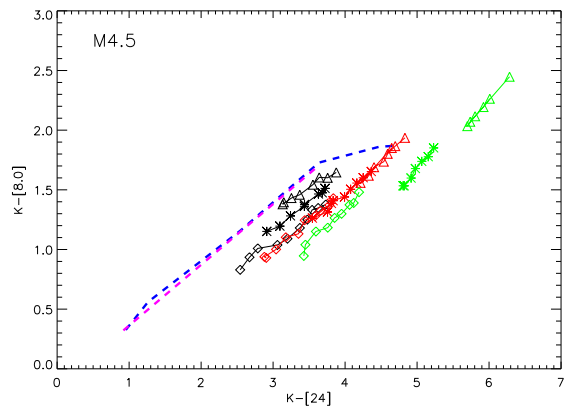}
\caption[]{Model evolutionary tracks for finite thickness
  discs surrounding an M4.5 star: note the different scale cf Figure 1. 
  Black, red and green symbols denote 
  optically thick models where b = 1, 1.13 and 1.25 
  respectively. Squares, asterisks and triangles denote  
  optically thick models where $H_D$ = 0.1, 0.5 and 1.0
 (see text
for definitions of b and $H_D$).
 In each case, the succession of points
  with the same symbol type and colour represent a sequence of
in each case at the point that the disc is viewed sufficiently
close to edge-on that the star itself is obscured ($A_v > 10$).
The dashed lines
  show the tracks for a uniformly draining model where $H_D =1$, b=1.13
  (blue = face on; magenta = close to edge on (cos i = 0.95 and 0.05)), with
the latter model only being plotted at periods when the central star is
optically visible.}
\label{f:f2}
\end{center}
\end{figure}

  We now repeat the above experiments (i.e we calculate the evolution in the
two colour plane of discs at a variety of inclinations which clear either
by uniform draining or by the growth of an inner hole) but now consider the 
case of discs where the dust has a finite thickness. An important difference 
from the razor thin case is that in  this case 
the line of sight to the star is now blocked at high inclinations
by the
disc and the K band flux is strongly reduced. We however do not plot any models
whose inclination is such that the extinction of the star by the disc
exceeds $A_v = 10$, since such objects would not appear in samples of
optically selected T Tauri stars. 

 Figure 2 plots a sequence of models at various inclinations (i.e. from
face-on to the most tilted case for which $A_v < 10$) for a range of
assumptions about the vertical structure of the disc. The vertical structure
of the dust disc is parameterised by two quantities that control
the degree of flaring and dust settling: $b$ is the parameter
that controls the radial variation of the disc scale height (i.e.
$H \propto R^b$). A disc with $b=1$ has a constant opening angle
whereas models with $b > 1$ are flared. In addition, a further
parameter represents the ratio ($H_D$)
of the scale height in the dust to that in
the gas: $H_D = 1$ corresponds to the case of well mixed gas and dust whereas
$H_D < 1$ denotes cases where the dust is relatively settled towards the mid-plane.

  Figure 2 shows that a variation in b over the range $1-1.25$  and of
$H_D$ in the range $0.1-1$ produces (in  combination  with a range of disc inclinations),  optically
thick model discs with a range of colours: as expected, the reddest
discs are flared, unsettled, face-on structures. 
The bluest colours that are produced by an optically thick disc
correspond to where the disc is observed  close to edge-on but with the
extinction to the star still satisfying $A_v < 10$. We find that even
where the dust is significantly settled (i.e. $H_D=0.5$), the bluest optically
thick (uncleared) disc has a value of K - [8] of around 1.2. Only
models with $H_D \sim 0.1$ can reach K - [8] $<$ 1.2 and these
represent a very extreme case of settling. 

  When these optically thick flared discs are subject to uniform draining they
evolve (for $\Sigma \propto R^{-1}$) along the dashed lines shown. As long
as the disc is optically thick in the optical, the models are still redder
(for a given surface density normalisation) than the flat disc models, since
the geometry of the optical photosphere sets the temperature distribution
in the disc. It is only at the point that the disc becomes optically
thin also to the star's optical radiation that disc geometry no longer
affects the disc colours. What is notable, however, is that (although the
flared disc models are 
redder than the flat discs in both colours for a given surface density profile) 
both flared and flat draining models lie along the same
trajectory in colour space. {\footnote{ The magenta line represents the evolution
of a uniformly draining flared disc viewed at large inclination angle
(cos i = 0.05) which appears in the plot only at the point that the
optical extinction to the star has fallen to $A_v < 10$. Because this
system is viewed almost edge-on, this point is only reached once the
disc has drained to very low surface densities: adopting the same disc
draining model as described above, this point is reached at very late times
($\sim 10 $ Gyr) , thus emphasising yet more the necessity of 
some accelerated draining at late times. It is notable,
leaving the timescale issue aside, that even such an extreme disc evolves
along the common draining locus once the central star becomes optically
visible}}. We have therefore shown that there exists a well defined draining
sequence such that  discs with a wide range of vertical structure and
inclinations  lie
along this trajectory.

The tracks for flared discs of various inclinations that
are subject to clearing through the progressive growth of inner holes
are qualitatively similar to the tracks with inner holes 
for flat discs except that, as expected, they are redder at K - [24].

\subsubsection{Summary}

\begin{figure*}
\begin{center}
\includegraphics[width=16cm]{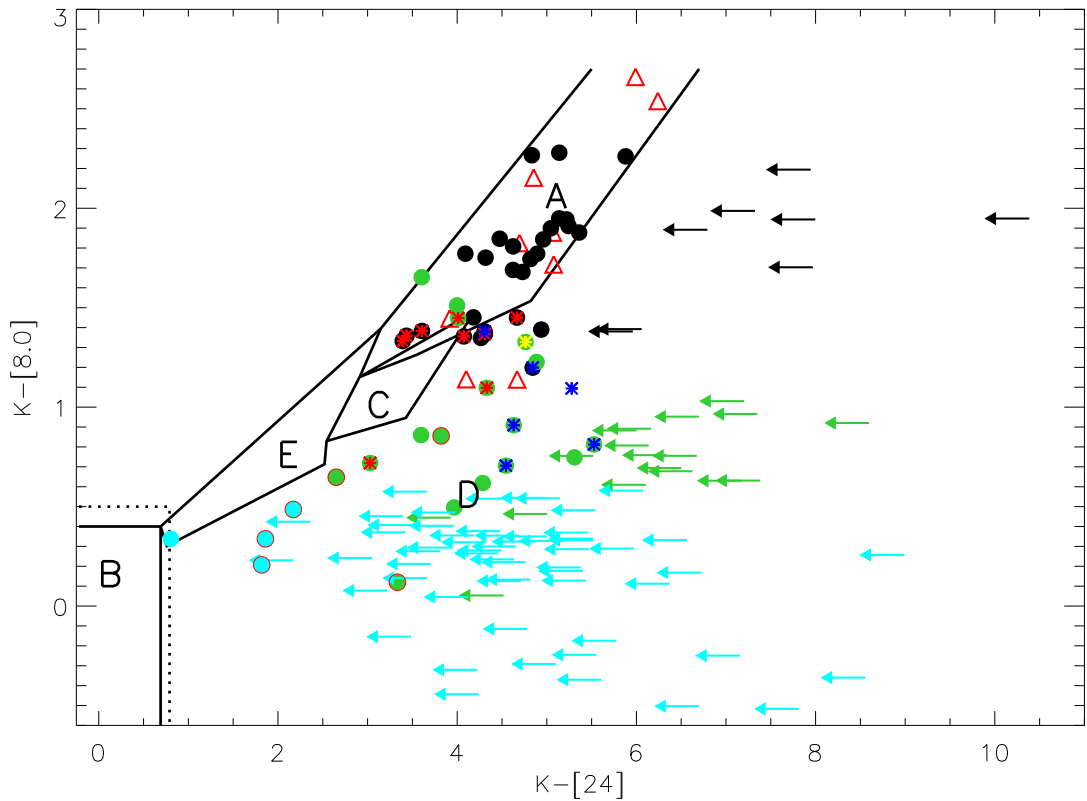}
\caption[]{Two colour plot showing the location of the M3 to M5
  sources in IC348 from the catalogues of Muench et al (2007) and Lada
  et al (2006) in the updated version of Currie \& Kenyon
  (2009). See Section 2.2 for details about the colour-codes and
  symbols. Five distinct regions have been labeled on this diagram;
  A: optically thick reprocessing discs; B: pure stellar photospheres;
  C: ultra-settled discs viewed close to edge on; D: inner hole
  sources; E: uniformly draining discs.    }
\label{f:f23}
\end{center}
\end{figure*}

  In Figure 3 we use the models computed above in order to classify 
various regions of the two colour plane. Regions A and B  correspond to
optically thick discs and pure stellar photospheres. The blueward extent
of region A  depends on the disc geometry: however even unflared discs 
(i.e. with
constant opening angle: $b=1$) have K - [8] values 
$> 1.2$  if the star is optically visible, unless the degree of
dust settling is  extreme (i.e. the case of $H_D = 0.1$).
The redward extent of region B  depends on the spectral type: the solid 
lines represent the
empirically derived infrared colours
of an M5 star presented in the Appendix of  Luhman et al (2010).
(K - [8] = 0.46, K - [24] = 0.69). The dotted line encompasses our estimate of the uncertainties
in this limit (see Section 3).

  Systems which lie outside these regions are `partially cleared discs',
apart from the small wedge of colour space (region C) which corresponds
to `ultra-settled' discs viewed close to edge-on. (`Ultra-settled' denotes
the case that the ratio of dust to gas scale height $H_D < 0.1$). Region D contains
objects that are compatible with being sources with cleared inner holes.
The strip E is the `draining locus', which corresponds to the
trajectory of discs in which the column density of dust is successively
reduced by a (spatially constant) factor. 

  Below we  use this plot in order to classify the disc bearing
properties of late type stars in IC 348.

\subsection{Matching to sources in IC 348}

\begin{figure*}
\begin{center}
\begin{minipage}[]{16.4cm}
\includegraphics[width=5.4cm]{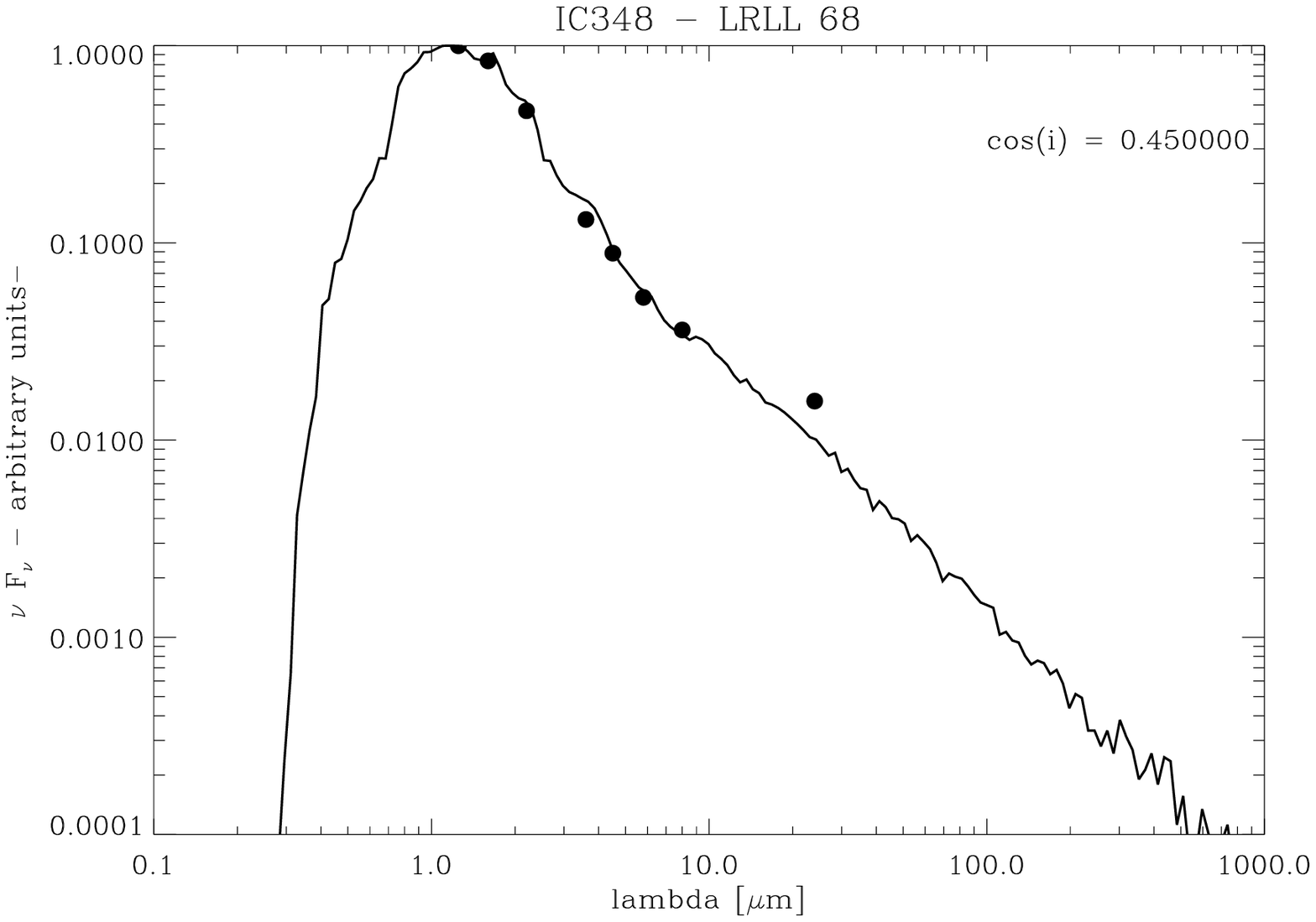}
\includegraphics[width=5.4cm]{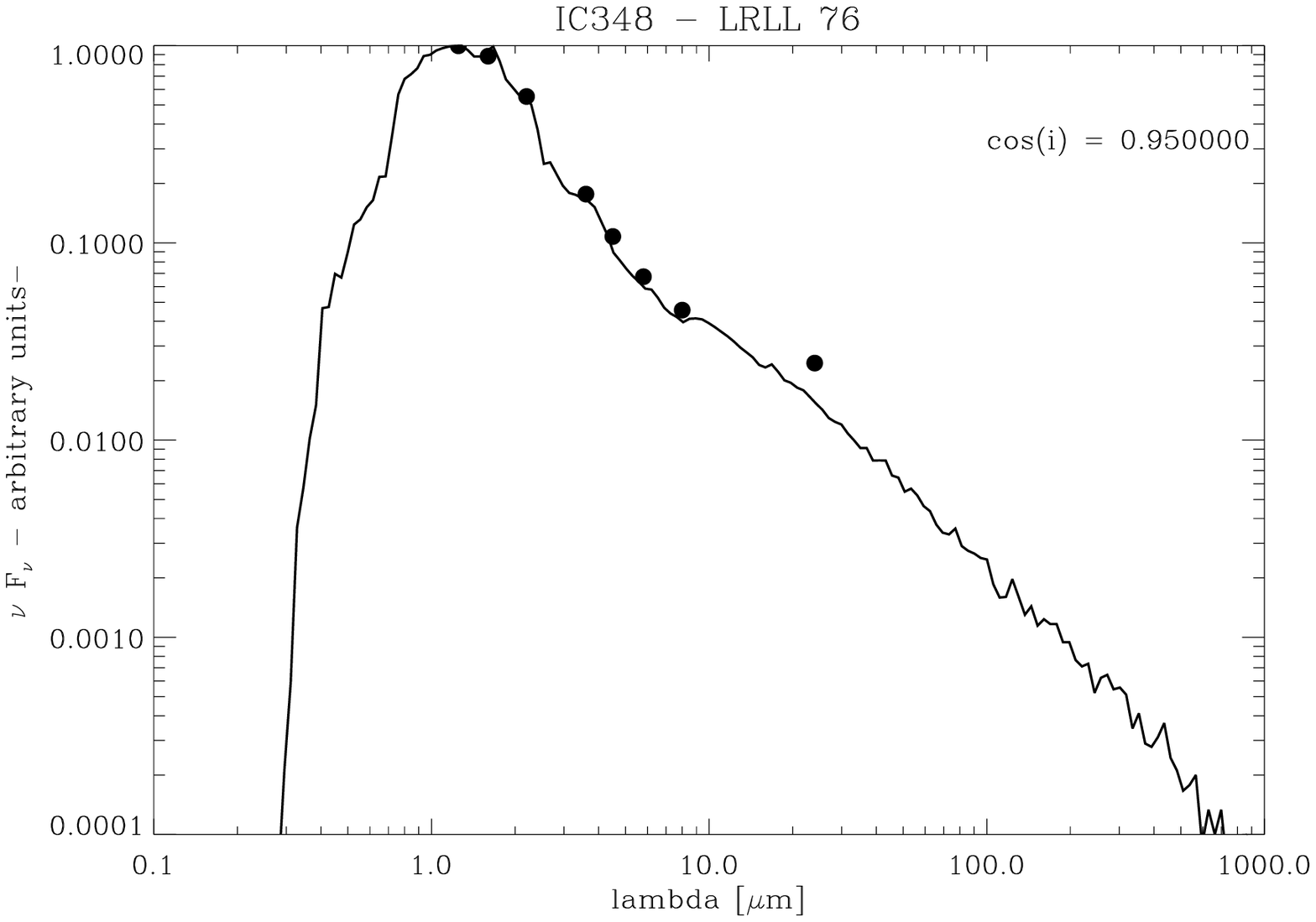}
\includegraphics[width=5.4cm]{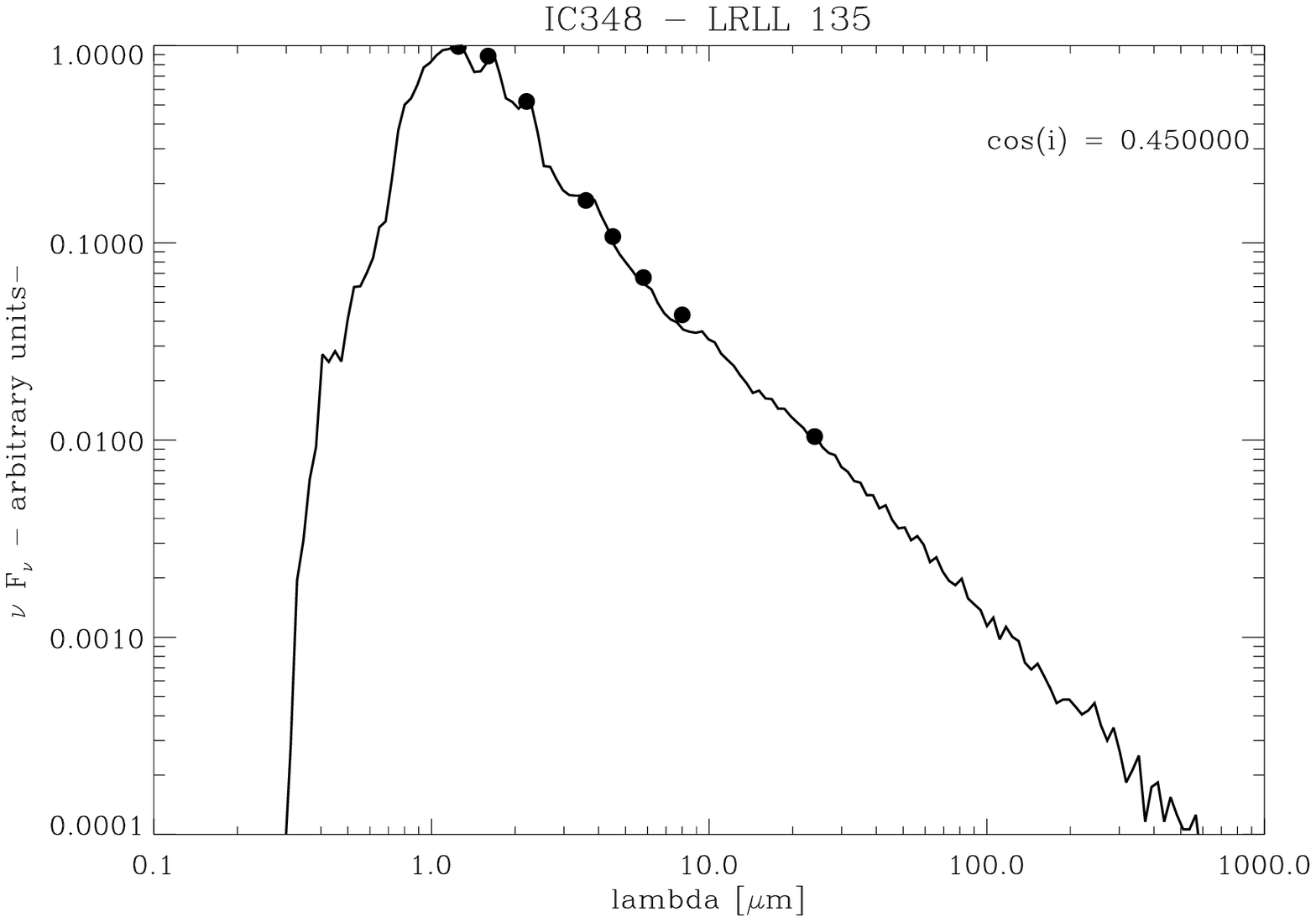}
\includegraphics[width=5.4cm]{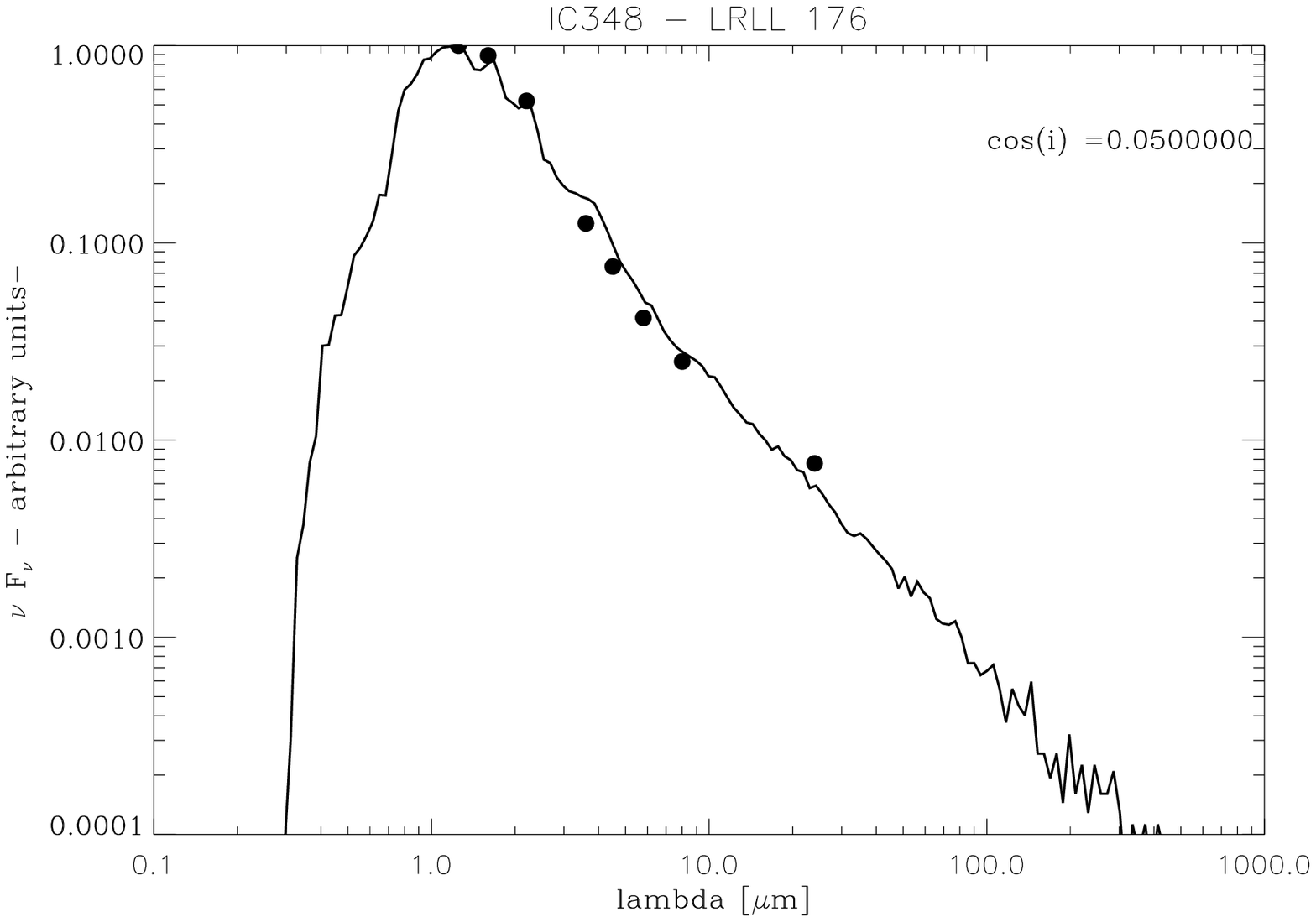}
\includegraphics[width=5.4cm]{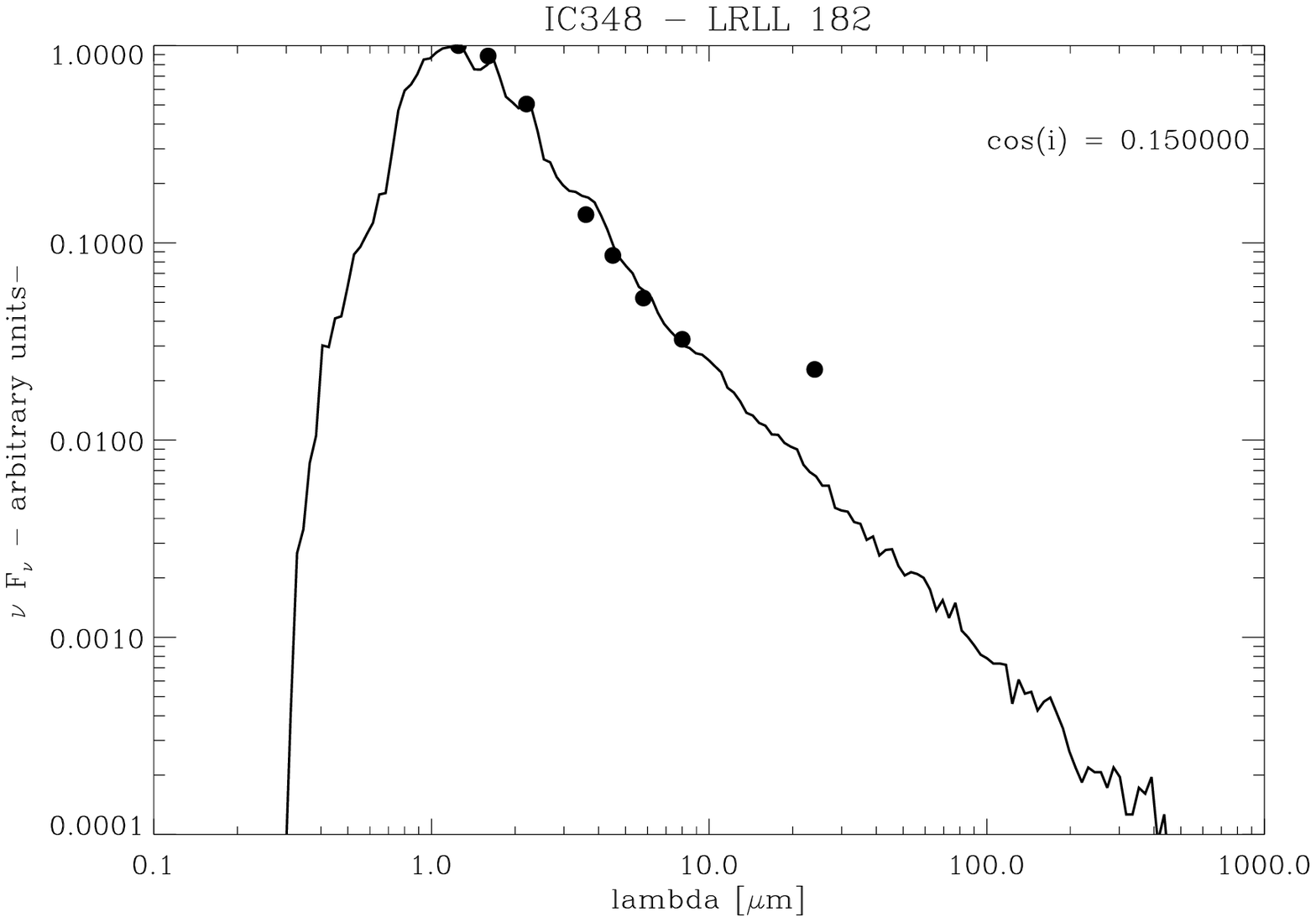}
\includegraphics[width=5.4cm]{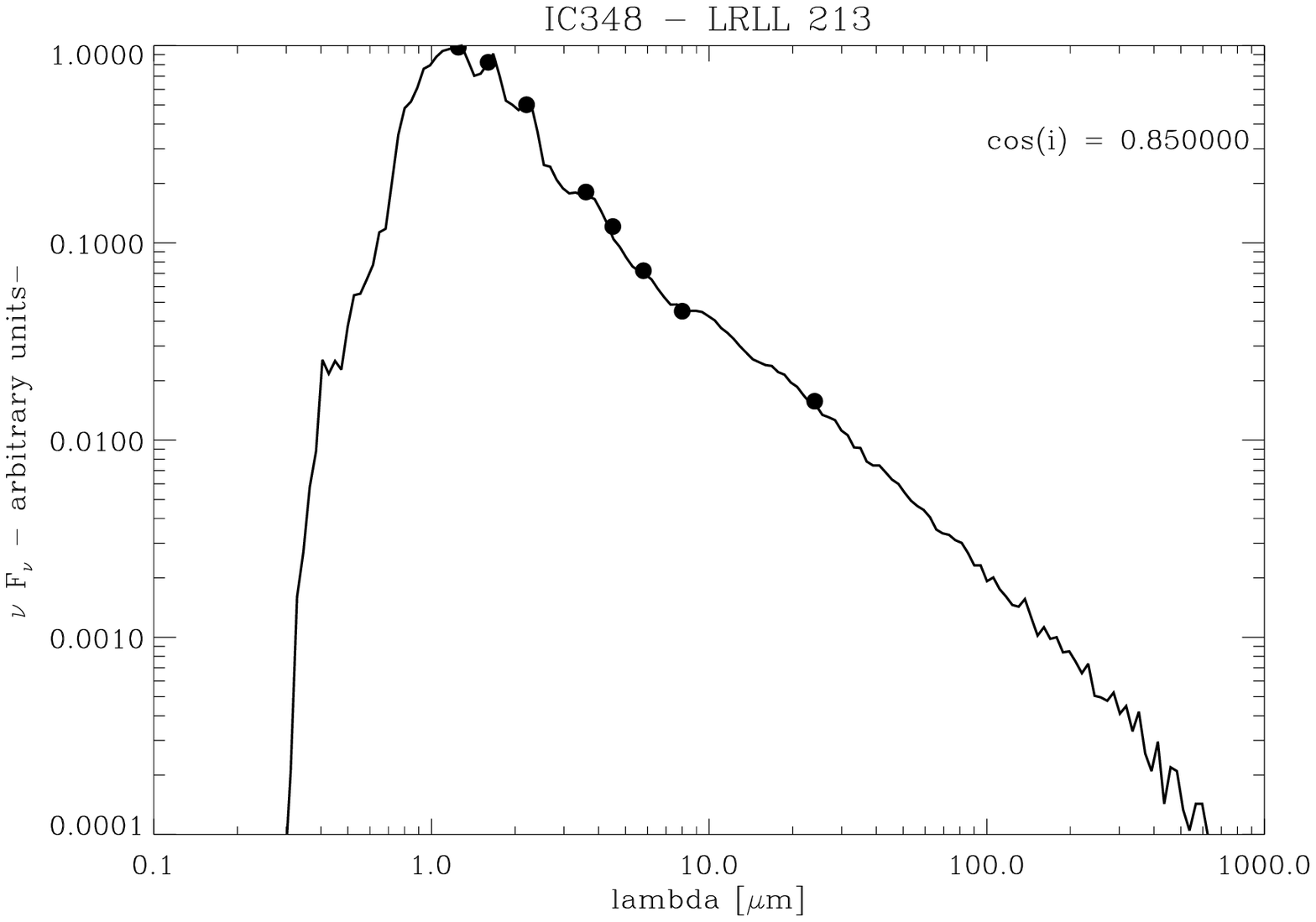}
\includegraphics[width=5.4cm]{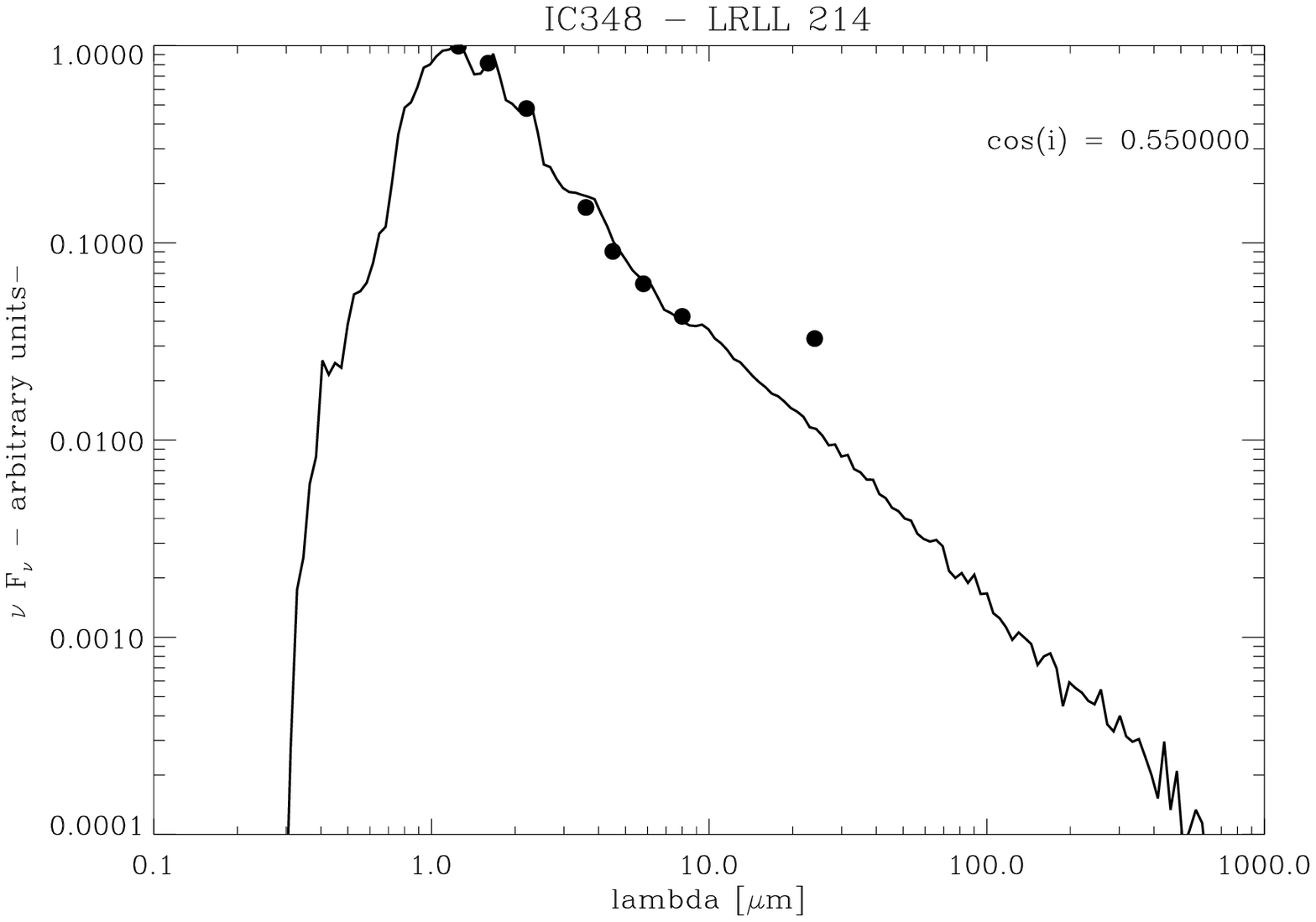}
\includegraphics[width=5.4cm]{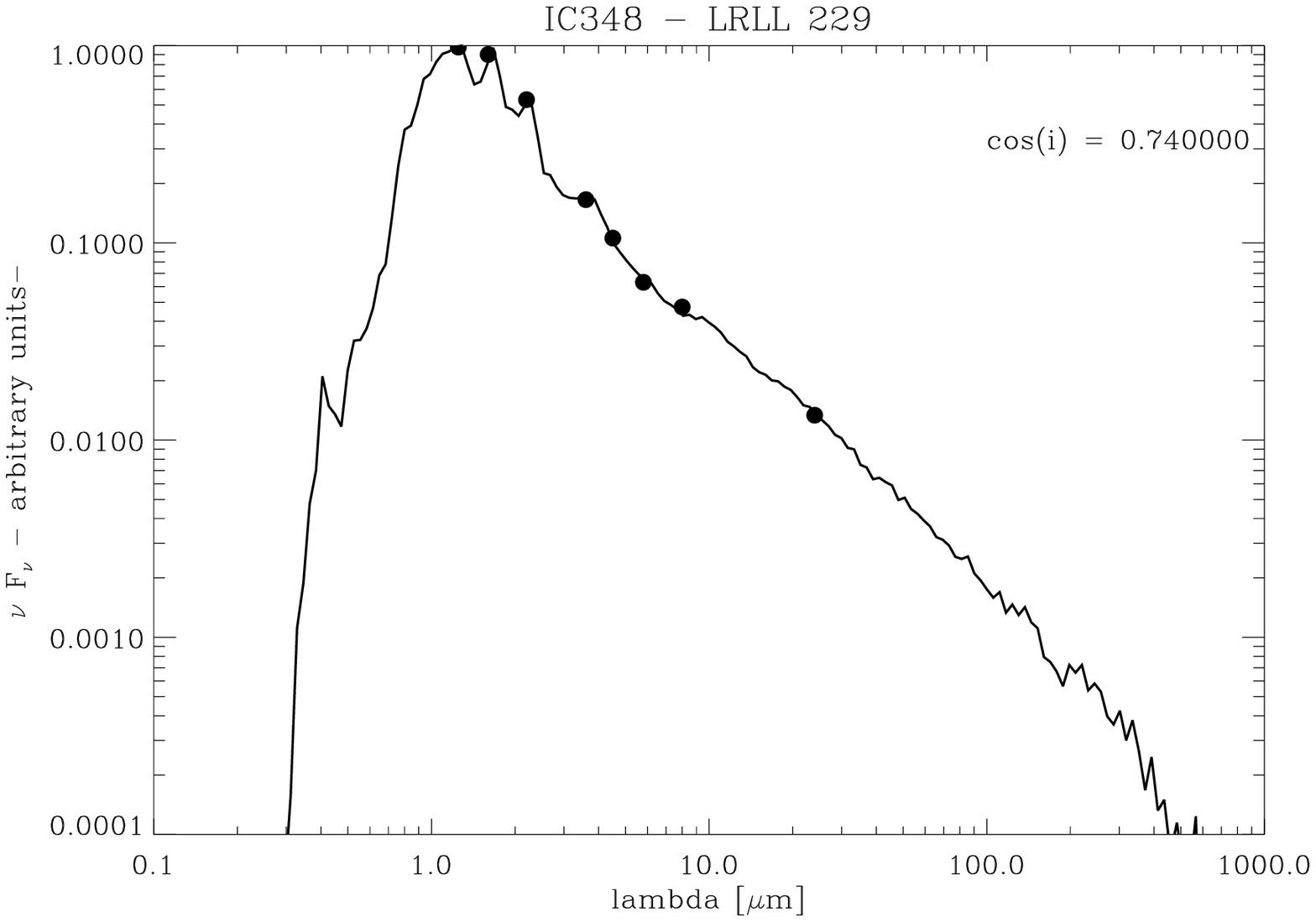}
\includegraphics[width=5.4cm]{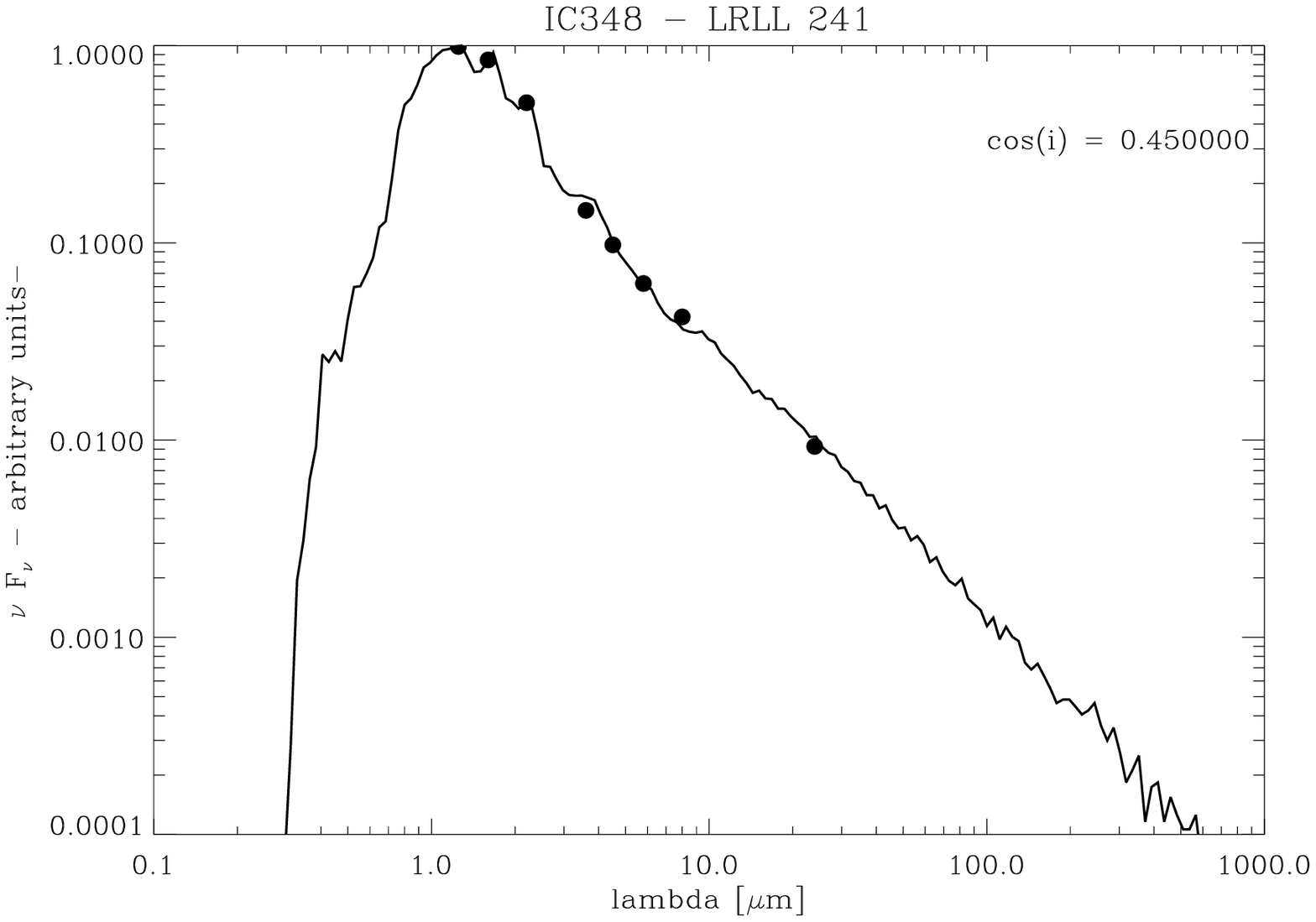}
\end{minipage}
\caption[]{Dereddened SEDs of the weak excess sources identified by
  Muzerolle et al (2010) in IC348 compared do model SEDs for optically
  thick flat reprocessing discs. In each case the best fit inclination
is indicated; such flat disc models are good fits in $5/9$ cases. Of
the remaining $4$ sources, three 
(LLRL 68,76,214) would require more flared disc structures whereas
LLRL 182 is suggestive of a cleared inner hole.}
\label{f:f4}
\end{center}
\end{figure*}

  We now turn to the question of how we should interpret the various
categories of partially cleared discs classified by previous authors.
Since many star forming regions (notably Taurus) contain relatively
few discs that occupy the partially cleared region, we here 
select a cluster in which it has been claimed that there is a
relatively large population of partially cleared discs (see 
Currie \& Kenyon 2010, Luhman et al 2010 for contrasting views
on this issue). 
We present a statistical analysis of IC 348 in Section 3 below
and here restrict ourselves to some exemplary cases from this cluster.
Figure 3 combines the classification  derived above with Spitzer colours of 
the stars in IC 348 in the spectral type range  M3 to M5,  dereddened
using the $A_v$ values of Lada et al (2006) and Muench et al (2007) and
the reddening law of Rieke \& Lebofsky (1985).  We have, where available,
revised the 
photometry obtained by Lada et al (2006) in accordance with the
more sensitive measurements (and MIPS photometry) obtained by
Currie \& Kenyon (2009).
 The black symbols
represent discs classified as (optically) thick discs by Lada et al (2006),
the green symbols are the `anaemic discs' of Lada et al (2006) and the blue
dots are systems classified as photospheres (on the basis of IRAC data
alone). The red triangles are the sources observed by Muench et al (2007).
The horizontal arrows represent the sources for which only
upper limits are available at $24 \mu$m which are unfortunately
quite numerous in this cluster owing to the bright background emission. 
The red, yellow and blue asterisks denote
sources that have variously been classified as `weak excess', `warm
excess' and `classical transition' (i.e. inner hole) sources by
Muzerolle et al (2010). The sources with an outer red circle
are those classified as `homologously depleted' by Currie \&
Kenyon (2009).

  It is immediately obvious from this plot that the various
classification schemes overlap and that we have sources that are
associated with a variety of designations. Thus there are red asterisks
with black borders (objects counted as weak excess by Muzerolle et al
2010, but as `thick discs' by Lada et al 2006) which have very
similar colours to other sources that were counted as `anaemic' by  
Lada et al (2006). Other anaemic sources (those with red border)
are described as `homologously depleted' by Currie \& Kenyon (2009).
Here we  draw attention to three  aspects of this plot:

 i) Almost all the `weak excess' discs of Muzerolle et al (2010)
lie close to the upper end of the line FE in Figure 1, implying that they are   
describable
as  {\it razor thin  optically thick discs} at a range of (relatively
face on) inclinations (i $< 60 \deg$). 
In order to check this,
we have fitted these sources as such discs in which the only
free parameter is the source inclination; the results
of this exercise are shown in Figure 4 and demonstrate excellent
agreement in 5/9 cases (namely sources LRLL135, LRLL176, LRLL213, LRLL229 and
LRLL241).  
Note that this is not a unique interpretation:
these same sources could also be fit by finite thickness (though rather
settled) optically thick
discs at somewhat larger inclinations (see Figure 2).   
The other 4 models have too much
flux at $24 \mu$m to be produced by a completely flat model; in three  of these
(sources LLRL 68, 76 and 214) the colours are compatible with a finite thickness
(but settled) optically thick disc (Figure 2). Only one source (LLRL 182)
has colours suggestive of a cleared
inner hole.  
Thus whereas Muzerolle et al (2010)
speculated that these weak excess models are in a state of partial
clearing, we demonstrate that they can mostly be fit by optically thick discs
in which the dust is somewhat settled towards the mid-plane. 
The sizeable population of sources
in this region of the two colour diagram is a demonstration that
disc clearing is preceded by  a significant period where the disc
is optically thick and with the dust well settled towards the mid-plane.
A similar conclusion concerning the prevalence of settled discs
in late type stars was reached by Pascucci et al (2003),
Apai  et al (2004) and  Allers et al (2006). 

 ii)  There are six   sources classified as `homologously depleted'
by Currie \& Kenyon (2009) in Figure 3 (red outer circles). 
The positions of
these sources in the two colour diagram are  suggestive of an inner hole and in
Figure 4, we illustrate  inner hole models for two  of these
sources (sources LLRL 124 and 261, located respectively at [K-24,K-8] of
[$2.17, 0.49$] and [$2.65,0.65$]). Amongst the many degeneracies that afflict dust SED modeling
of discs, we highlight that there is a degeneracy between the disc
inclination and inner hole size, as shown in Figure 5 where the two
sources are equally well matched with flat discs with [8] or 20
AU holes of different inclination. 

iii) Over all we see that among the partially cleared discs (see
Figure 3),
there are {\it no detected sources} close to the draining locus.
The interpretation of this result is complicated by the large
number of $24 \mu$m upper limits in IC 348. These upper limits are
actually higher (in K - [24]) than the detected sources; the simplest
explanation is that higher sensitivity observations would yield
K - [24] colours in the same region of the two colour diagram as
the detected sources (i.e. would place them in the inner hole
region D). We cannot of course rule out the alternative possibility that they
are actually much bluer in K - [24] than the inner hole sources (i.e.
that they would lie on the draining locus). There is no obvious
reason, however, why sources that are intrinsically blue in K - [24]
should be systematically
associated with large upper limits on K - [24] colour.   
However, the only definitive statement that we can make here is that there
is no evidence for any sources being close to the draining locus
in this cluster, despite the large number of sources in the partially
cleared region. 

\section{The statistics of disc clearing in IC 348 \& Taurus}
\begin{figure*}
\begin{center}
\begin{minipage}[]{10.8cm}
\includegraphics[width=5.4cm]{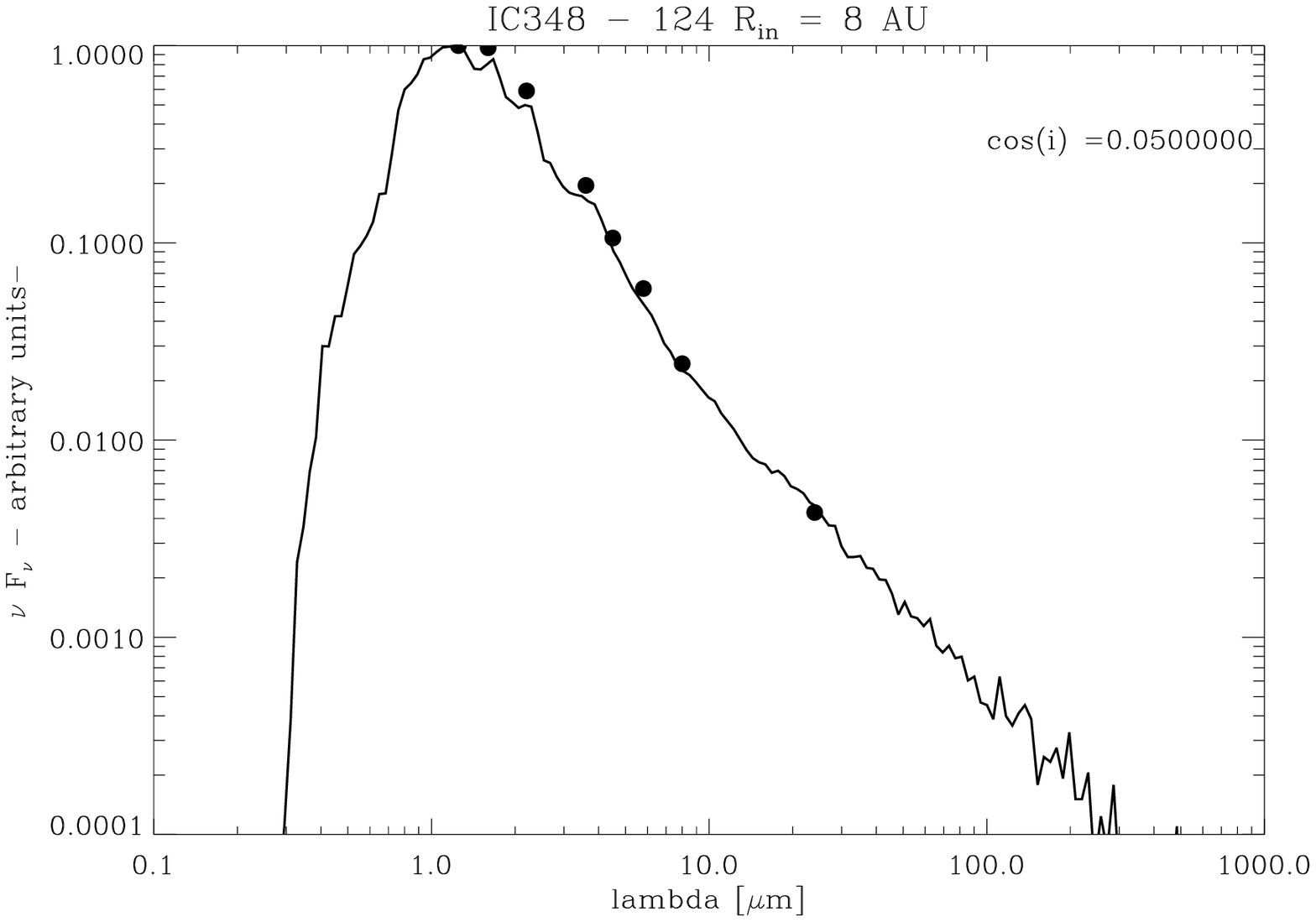}
\includegraphics[width=5.4cm]{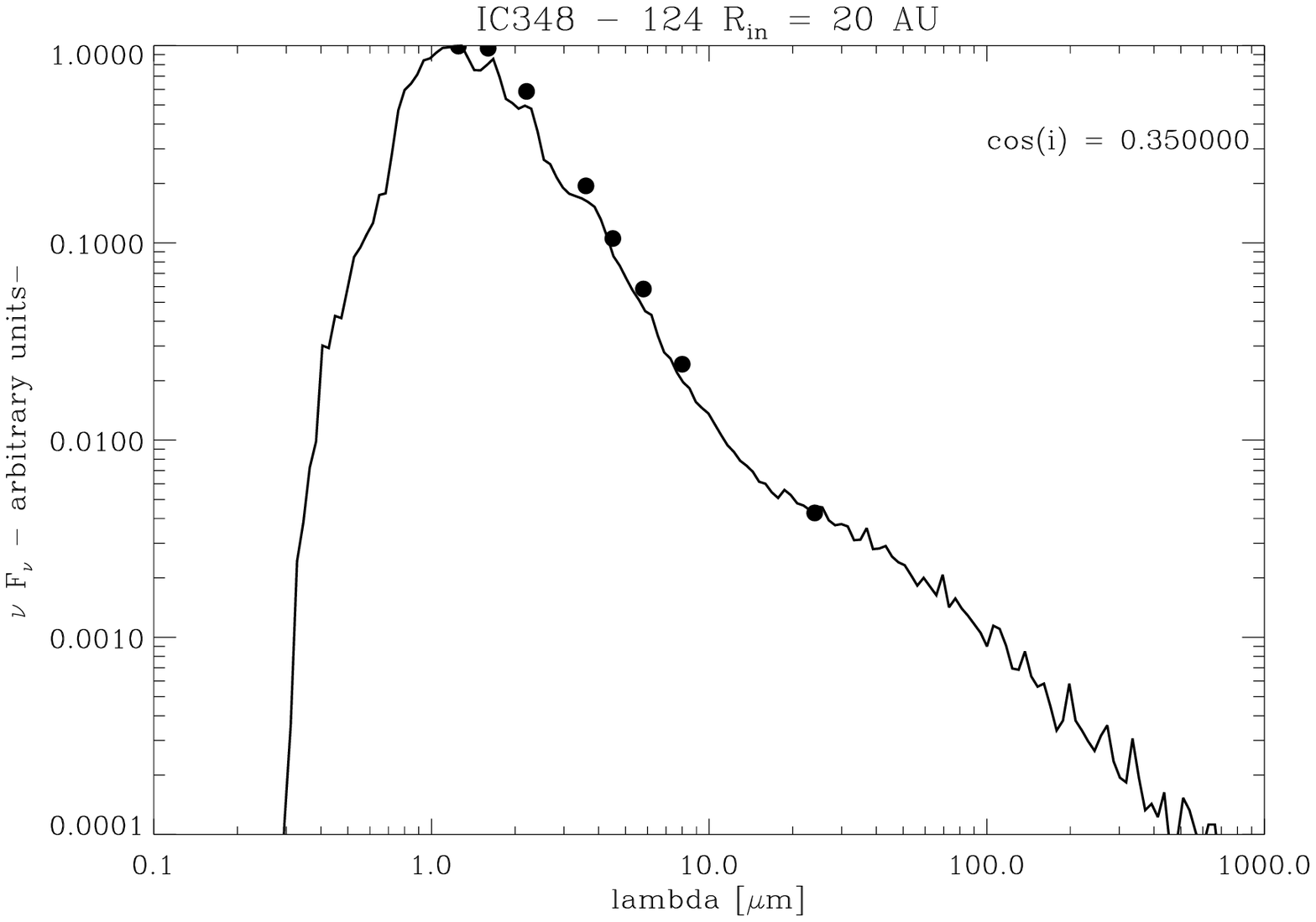}
\includegraphics[width=5.4cm]{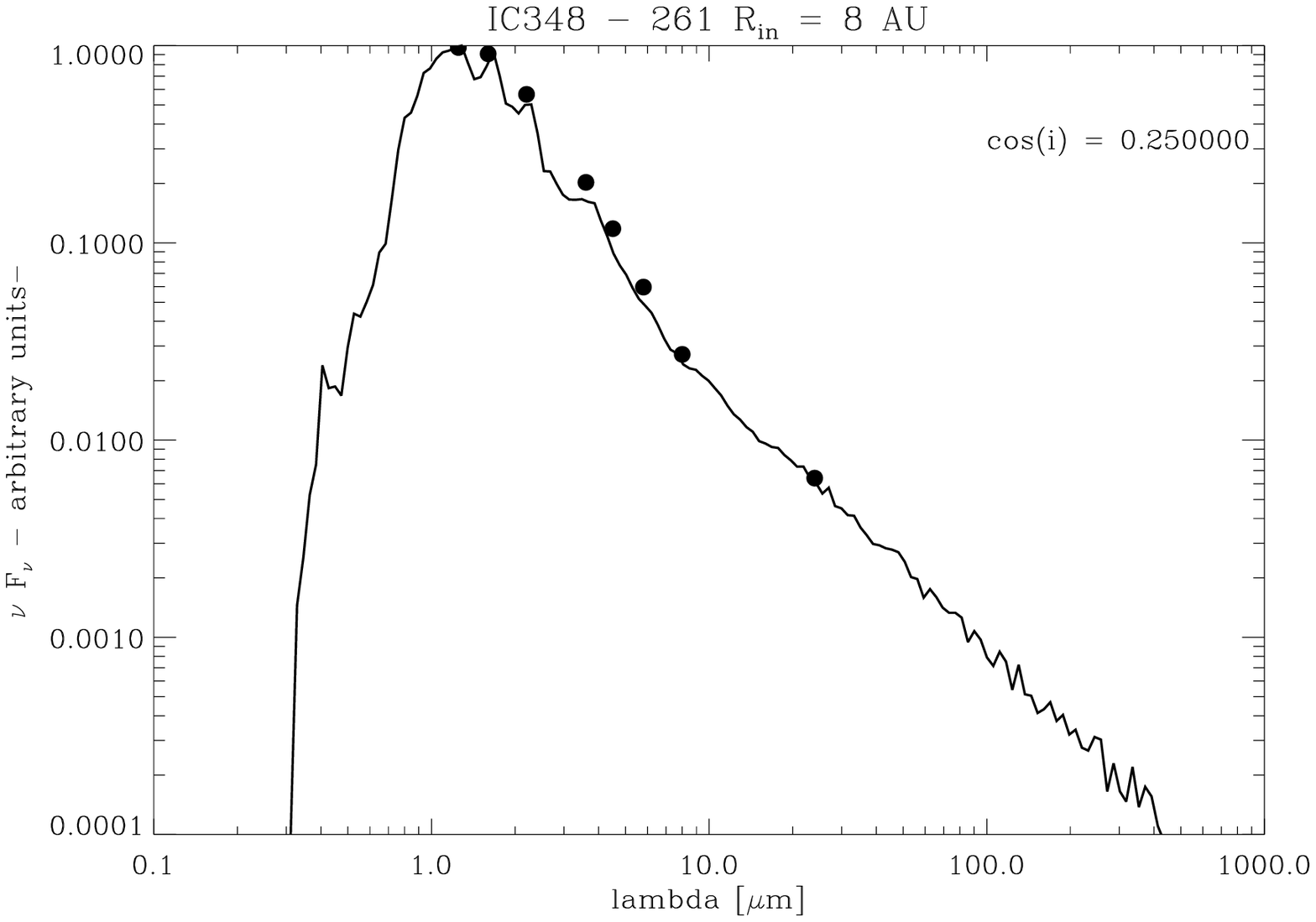}
\includegraphics[width=5.4cm]{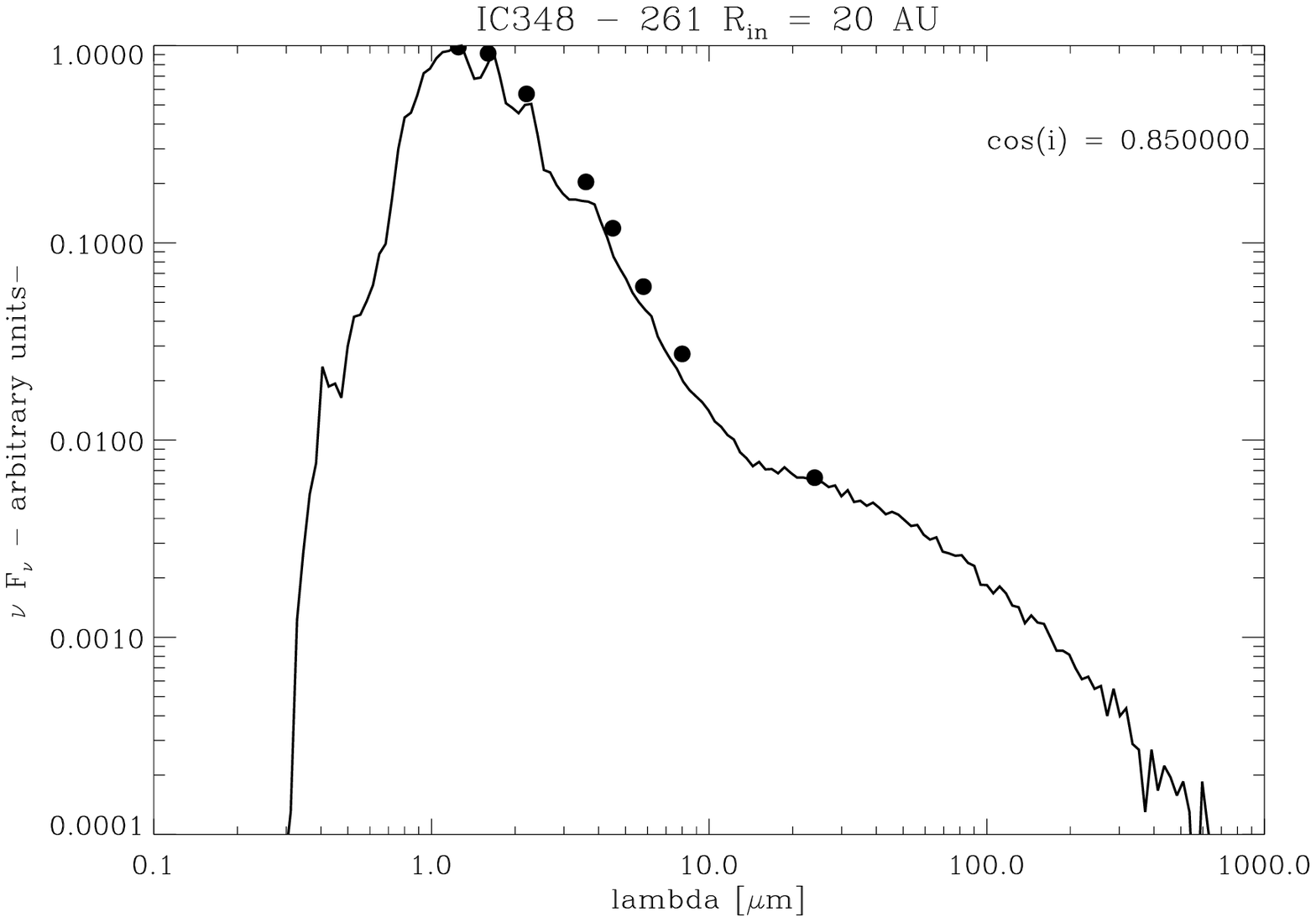}
\end{minipage}
\caption[]{Dereddened SEDs for two sources in IC 348, identified as 'homologously
depleted' by CK09, compared to model SEDs of flat optically thick discs
with inner holes of [8]~AU (left panels) and 20~AU (right panels).}
\label{f:f5}
\end{center}
\end{figure*}

  The discussion in 2.2 allowed us to divide the infrared two colour plane
into distinct regions: broadly speaking, the K - [8] colour {\it alone}
is sufficient to determine whether a disc is in a partially cleared
state, whereas, for systems in this condition, the $24 \mu$m flux is
required in order to distinguish the {\it mode} of disc clearing
(i.e. uniform draining versus inner hole growth). There are two regions
however where the identification by K - [8] colour alone is not completely
unambiguous.  
Firstly, 
there is an overlap in K - [8] colours (in the range 0.8-1.2) between partially
cleared discs and ultra-settled edge-on discs (i.e. region C in
Figure 3).  Secondly there is some uncertainty (of the order of a few
tenths of a magnitude) in the red boundary of the K - [8] colour 
of pure stellar photospheres. 
  
 Bearing these caveats in mind, we can  use the distribution of K - [8] colours in order to
look at the statistics of the incidence of clearing discs. Luhman et al
(2010) presented
such an analysis (in terms of the spectral slope between $2$ and $8 \mu$m)
for a number of clusters and concluded that  - for stars of given
spectral type - the distribution of spectral
slopes among disc bearing stars in IC 348 was 
indistinguishable from that in Taurus.  Since Taurus is the original
region in which the small number of transition objects was first noted,
it would then follow that such objects are comparably rare
in IC 348. 

 This conclusion is at first
sight surprising when one compares Figure 3 with the corresponding
two colour plot for stars in the same range of spectral type in Taurus,
as there would, to the eye, appear to be a much more pronounced
gap in the Taurus data (Figure 12 of Luhman et al). 
It is however important to note that Luhman based his comparison
between Taurus and IC 348 only 
on those sources that were detected at $24 \mu$m:
from Figure 3 one sees that  there are many upper limits at $24 \mu$m in the
IC 348 two colour plot and that these upper limits 
are particularly
numerous among  sources that are blue in K - [8]: i.e. among the
sources undergoing disc clearing. Although the fraction of systems that
are undetected at $24 \mu$m is much lower in Taurus, it is again
the case that those systems that are not detected at $24 \mu$m tend to
be  those
with blue K - [8] colours. The systematic omission of sources
that have only upper limits at $24 \mu$m will thus skew any comparison
of the distribution of discs as a function of K - [8] colour.

  With this in mind, we re-analyse the distribution of  K - [8] colours
of the de-reddened SEDs of M3-M5 stars in IC 348 and Taurus.{\footnote{
Luhman et al list values of the spectral slope $\alpha$, where 
\begin{equation}
\alpha = {{d {\rm{log}} (\lambda F_\lambda)}\over{{\rm{d log}} (\lambda)}}
\end{equation}
\noindent over the spectral region $2.2-8 \mu$m and we convert these
values to K - [8] colours using
K - $[8] = -2.5 {\rm log}_{10} \biggl( 0.275^{\alpha}/37.8\biggr)$
}}
 We include all sources (irrespective
of whether they are detected at $24 \mu$m)  and have $83$ such sources
in Taurus (with $\alpha$ values taken from Table 7 of Luhman et al)
and $144$ sources in IC 348 with K - [8] 
values from the photometry sources listed in Section 2 above. 

\begin{figure}
\begin{center}
\includegraphics[width=8cm]{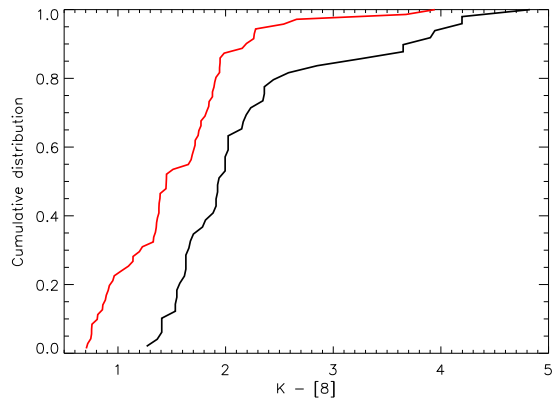}
\caption[]{Cumulative distribution of K - [8] colours for the M3 to M5 stars in Taurus (black line) and IC348 (red line), including only those sources
which have K - [8] $>$ 0.7 and where we can be confident that there is some
form of circumstellar emission.}
\label{f:f6}
\end{center}
\end{figure}

 If
we simply compare the distributions of K - [8] colour in the two clusters
with a Kolmogorov Smirnoff (KS) test, we obtain the result that
the two distributions are overwhelmingly different (KS probability
of $\sim 10^{-4}$ of these datasets being drawn from the same
population).
This highly significant difference however simply reflects the fact
that IC 348 
contains many more
systems that have completely cleared their discs. What we wish to test
however is whether, {\it within the population of objects that retain
some form of discs} there is a higher fraction of partially cleared discs
in the case of IC 348.

 At this point we come up against the problem of defining the
red limit in K - [8] colour for purely photospheric emission.
The combination of photometric errors, the finite range of spectral types
in our sub-sample  and uncertainties in the empirical calibration 
of spectral type against photospheric colour means that it is difficult
to propose a robust criterion for the K - [8] colour of a discless system.
The two colour diagram of detected sources in Taurus illustrates this
problem: even within the spectral type range of M3 to M5, there are
objects with K - [8] colour of around 0.5 with very different K - [24]
colours, suggesting that objects close to this K - [8] colour are a mixture of 
photospheres and  systems with discs. In order to avoid this ambiguity,
we therefore restrict our comparison of K - [8] colour distributions
to those objects that are redder in K - [8] than 0.7. In this way, we
can be confident that all the objects in our sample (in each cluster) have
some form of circumstellar emission; inevitably, in the process we omit
a few  bluer objects that have some residual disc. But for our 
present purposes,
all that matters is that we have samples in the two regions that contain
all stars of a given spectral type with K - [8] $>0.7$  and that this selection
ensures that all the objects in our samples show some form of circumstellar
emission (i.e. we are comparing disc systems - whether partially cleared
or not - in both regions).   

 In Figure 5 we plot cumulative distributions
for K - [8] colour for the two regions (with IC 348 being the red curve that 
attains a value of unity at lower K - [8]), assuming a cut-off in
 K - [8] colour of 
$0.7$ discussed above. 
This choice highlights the dearth of Taurus
objects with K - [8] in the range $0.7-1.2$
i.e. the range of colours where discs are either partially
cleared or ultra-settled (note that there {\it are} several partially
cleared (inner hole sources) in Taurus but  these have bluer
K - [8] colours than those sources in Figure 6 and we would not be able
to identify these sources as such in the absence of 
$24 \mu$m data). By contrast,
around $30 \%$ of disc bearing objects in IC 348 have spectral slopes
in the colour range $0.7 < K - [8] < 1.2$ in which discs are either
partially cleared or ultra-settled. The KS probability of these distributions being the
same is $10^{-5}$.

  We thus conclude 
that the distribution of K - [8] colours in IC 348
and in Taurus are very different once one includes all objects (regardless
of whether they are detected at $24 \mu$m). We can estimate 
the fraction of disc bearing systems that are partially cleared
or ultra-settled by adding all the sources in the colour range
$0.7 < K - [8] < 1.2$ (regardless of whether they are detected at 
$24 \mu$m or not) to all bluer objects with a measured
excess at  $24 \mu$m.  This yields a total of $2 + 6$  in Taurus
and $23 + [8]$ in IC 348, compared with a total number of disc bearing
objects of 55 and [8]1 in the two clusters respectively. This corresponds
to a fraction of discs that are partially cleared  of
$11 \%$ in Taurus and $38 \%$ in IC 348.  This is however a lower limit since
it is possible that some fraction of the objects with photospheric
K - [8] colours but with $24 \mu$m upper limits that are above photospheric
levels (these objects numbering $21$ in Taurus and nearly $70$ in IC 348)
could turn out to have residual outer discs.

\section{Conclusions}

  Our modeling has allowed us to propose a classification scheme
for late type T Tauri stars based on their locations in the K - [8], K
- [24]
two colour plane (Figure 3). We highlight the fact that this diagram
reveals a locus for discs that clear through uniform draining (i.e.
reduction in surface density by a constant factor) which is quite
distinct (much bluer in K - [24] for given K - [8]) than the region occupied
by discs with cleared inner holes. This is because a uniformly draining disc 
becomes optically thin first at large radii and longer wavelengths.
Figure 3 also contains data for IC 348
where symbol type and colour illustrates the overlapping designations
proposed by previous authors. 

  We find that a few of the anaemic discs of Lada et al (2006) and most
of the discs described as `weak excess' by Muzerolle et al (2010) are
compatible with rather flat/settled optically thick dust discs. Their 
abundance in this cluster is therefore an indication of dust settling
rather than disc clearing. 

  There are however a large number of sources (at least $38 \%$ of systems
with evidence for circumstellar emission in IC 348) with evidence for partial
clearing of the disc. We have shown that (provided one does not omit
sources that are undetected at $24 \mu$m) the fraction of discs in a
partially cleared state is at least a factor of $3-4$  higher in IC 348 than
in Taurus (for stars in the same range of spectral type); the difference
in K - [8] colour distributions in the two clusters is highly statistically
significant (KS probability of $10^{-5}$ of being drawn from the same
distribution). IC 348 is apparently then a cluster with a more evolved
disc population, consistent also with the fact that it contains a larger
population of highly settled discs. 

  Figure 3 also allows us to distinguish between different categories
of partially cleared discs. All objects that are detected at $24 \mu$m
lie in the domain of the two colour diagram corresponding to discs with
cleared inner holes. It is unfortunate that, given the high nebula
background emission in IC 348, there are many sources that are
undetected at $24 \mu$m and for which we cannot therefore assess
whether they are inner hole sources or those undergoing uniform draining.
The maximum possible fraction of partially cleared sources which could
conceivably be uniform draining (on the extreme assumption that all
undetected sources are uniform draining) is around a half. We conclude
therefore that the dominant clearing channel in IC 348 is by inner hole
creation.  We also note that in  regions  
where the sensitivity at $24 \mu$m {\it would} allow sources to be
detected on the draining locus (Taurus, $\eta$ Cha, Chamaeleon, Upper Sco)
and where there are collectively around $20$ sources that are partially
cleared, none are located on the draining locus. This reinforces our view
that {\it inside out clearing is the dominant evolutionary path}.

  Finally we stress that although we have shown that the fraction of
disc bearing stars that show evidence of partial 
clearing is significantly higher
in IC 348 than in Taurus (by a factor $3-4$), this does
{\it not}  imply that the clearing timescale is necessarily
longer: indeed it would be surprising if the clearing timescale did vary
from region to region for stars of given spectral type. A significant
difference between IC 348 and Taurus is that in the former case the majority
of stars have already lost their discs, and therefore we have been  comparing 
the number of discs in transition with the minority of systems that still
possess discs. This fits well with the age difference between IC 348
(4-5~Myr) and Taurus (1.5 Myr).
If one instead compares the number of partially cleared discs
with the total sample of M3-M5 stars in both regions we obtain ratios  
of $31/144 = 21 \%$ in IC 348 compared with $8/83 = 10 \%$ in Taurus. 
Given the small numbers involved and the difficulties in categorising
individual marginal objects, especially given photometric uncertainties,
it cannot be considered that this is a significant difference. We thus
conclude - both in Taurus and IC 348 -  that a typical late type T Tauri  star
spends a comparably small fraction of its life in a state of disc clearing  
($10-20 \%$). Our results thus confirm that disc clearing is {\it rapid}
(a few times $10^5$ years) and from the inside out and further motivate the 
development of models that satisfy these criteria.  
\section{Acknowledgments}
BE acknowledges support from an STFC Advanced Fellowship. We thank
Nathan Mayne for helpful discussion with regards to relative
age-dating of clusters.


\begin{thebibliography}{}
\bibitem[]{} Alexander, R., Clarke, C., Pringle, 2006. MNRAS 369,229

\bibitem[]{} Allen, L., et al.\ 2007, Protostars and Planets V, 361

\bibitem[]{} Allers, K., Kessler-Silacci, J., Cieza, L., Jaffe, D., 2006. ApJ 644,364

\bibitem[]{} Apai, D., Pascucci, I., Sterzik, M., van der Bliek, N., Bouwman, J., Dullemond, C., Henning, T., 2004. A \& A L426,53

\bibitem[Armitage \& Hansen (1999)] {} Armitage, P. \& Hansen, B., 1999, Nature 402,633

\bibitem[]{}Clarke, C., Gendrin. A., Sotomayor, M, 2001. MNRAS 328,485

\bibitem[Currie \& Kenyon(2009)]{2009AJ....138..703C} Currie, T., \& Kenyon, S.~J.\ 2009, \aj, 138, 703 

\bibitem[Ercolano et al.(2008)]{2008ApJ...688..398E} Ercolano, B., Drake, 
J.~J., Raymond, J.~C., \& Clarke, C.~C.\ 2008b, \apj, 688, 398 

\bibitem[Ercolano et al.(2009)]{2009ApJ...699.1639E} Ercolano, B., Clarke, 
C.~J., \& Drake, J.~J.\ 2009, \apj, 699, 1639 

\bibitem[Ercolano 
\& Clarke(2010)]{2010MNRAS.402.2735E} Ercolano, B., \& Clarke, C.~J.\ 2010, \mnras, 402, 2735 

\bibitem[Ercolano et al.(2009)]{2009MNRAS.394L.141E} Ercolano, B., Clarke, 
C.~J., \& Robitaille, T.~P.\ 2009, \mnras, 394, L141 

\bibitem[]{} Evans, N.~J., et al. 2009, \apjs, 181, 321

\bibitem[Gorti 
\& Hollenbach(2009)]{2009ApJ...690.1539G} Gorti, U., \& Hollenbach, D.\ 2009, \apj, 690, 1539 

\bibitem[]{} Guieu, S., et al.\ 2009, \apj, 697, 787

\bibitem[]{} Gutermuth, R.~A.,Megeath, S.~T., Myers, P.~C., Allen, L.~E., Pipher, J.~L.,\& Fazio, G.~G.\ 2009, \apjs, 184, 18

\bibitem[Haisch et al.(2001)]{2001ApJ...553L.153H} Haisch, K.~E., Jr., 
Lada, E.~A., \& Lada, C.~J.\ 2001, ApJl, 553, L153 

\bibitem[Hartmann et al.(1998)]{1998ApJ...495..385H} Hartmann, L., Calvet, 
N., Gullbring, E., \& D'Alessio, P.\ 1998, \apj, 495, 385 

\bibitem[Kenyon 
\& Hartmann(1995)]{1995ApJS..101..117K} Kenyon, S.~J., \& Hartmann, L.\ 1995, \apjs, 101, 117 

\bibitem[]{} Koenig, X. P., Allen, L. E., Gutermuth, R. A., Hora, J. L., Brunt, C. M., 2008, \& Muzerolle, J., \apj, 688, 1142

\bibitem[Lada et al.(2006)]{2006AJ....131.1574L} Lada, C.~J., et al.\ 2006, 
\aj, 131, 1574 

\bibitem[Luhman et al.(2010)]{2010ApJS..186..111L} Luhman, K.~L., Allen, 
P.~R., Espaillat, C., Hartmann, L., \& Calvet, N.\ 2010, \apjs, 186, 111 

\bibitem[Mayne et al.(2007)]{2007MNRAS.375.1220M} Mayne, N.~J., Naylor, T., 
Littlefair, S.~P., Saunders, E.~S., 
\& Jeffries, R.~D.\ 2007, \mnras, 375, 1220 

\bibitem[Mayne 
\& Naylor(2008)]{2008MNRAS.386..261M} Mayne, N.~J., \& Naylor, T.\ 2008, \mnras, 386, 261 

\bibitem[Muench et al.(2007)]{2007AJ....134..411M} Muench, A.~A., Lada, 
C.~J., Luhman, K.~L., Muzerolle, J., \& Young, E.\ 2007, \aj, 134, 411 

\bibitem[Muzerolle et al.(2010)]{2010ApJ...708.1107M} Muzerolle, J., Allen, 
L.~E., Megeath, S.~T., Hern{\'a}ndez, J., 
\& Gutermuth, R.~A.\ 2010, \apj, 708, 1107 

\bibitem[Owen et al.(2010)]{2010MNRAS.401.1415O} Owen, J.~E., Ercolano, B., 
Clarke, C.~J., \& Alexander, R.~D.\ 2010, \mnras, 401, 1415 

\bibitem[]{} Pascucci, I., Apai, D., Henning, T., Dullemond, C., 2003. ApJ L590,111

\bibitem[]{} Rebull, L.~M., et al.\ 2010, \apjs, 186, 259

\bibitem[Whitney et al.(2003)]{2003ApJ...591.1049W} Whitney, B.~A., Wood, 
K., Bjorkman, J.~E., \& Wolff, M.~J.\ 2003, \apj, 591, 1049 

\bibitem[Whitney et al.(2003)]{2003ApJ...598.1079W} Whitney, B.~A., Wood, 
K., Bjorkman, J.~E., \& Cohen, M.\ 2003, \apj, 598, 1079 

\bibitem[Yasui et al.(2009)]{2009arXiv0908.4026Y} Yasui, C., Kobayashi, N., 
Tokunaga, A.~T., Saito, M., \& Tokoku, C.\ 2009, arXiv:0908.4026 

\end{thebibliography}
\end{document}